\documentstyle[12pt]{article}
\title{Some eigenstates for a model associated with solutions
       of tetrahedron equation.\\
       II.~A bit of algebraization}
\author{I.G.~Korepanov\\
\footnotesize $\matrix{ \cr
       \hbox{Chelyabinsk University of Technology}\cr
       \hbox{76 Lenin av., Chelyabinsk 454080, Russia}
       }$}
\date{February 1997}
\def\be{\begin{equation}}
\def\ee{\end{equation}}
\makeatletter
\long\def\@makecaption#1#2{\vskip 10\p@ \hbox to\hsize{\hfil#1\hfil}}
\def\cotanh{\mathop{\operator@font cotanh}\nolimits}
\makeatother
\newbox\ALH
\setbox\ALH=\hbox{\unitlength=1mm
 \begin{picture}(2.40,2.00)
 \put(1.20,1.00){\circle{2.00}}
 \put(1.20,1.00){\makebox(0,0)[cc]{.}}
 \end{picture}}
\newbox\ALHindex
\setbox\ALHindex=\hbox{\unitlength=1mm
 \begin{picture}(1.68,1.40)
 \put(0.84,0.70){\circle{1.40}}
 \put(0.84,0.70){\makebox(0,0)[cc]{.}}
 \end{picture}}
\def\alh{\mathchoice{\copy\ALH}{\copy\ALH}{\copy\ALHindex}{\copy\ALHindex}}
\begin{document}
\maketitle

\begin{abstract}
This paper adds two observations to the work solv-int/9701016
where some eigenstates for a model based on tetrahedron equation
have been constructed. The first observation is that there exists 
a more ``algebraic'' construction of one-particle states,
resembling the $1+1$-dimensional algebraic Bethe ansatz.
The second observation is that the strings introduced
in solv-int/9701016 are symmetries of a transfer matrix,
rather than just eigenstates.
\end{abstract}

\section*{Introduction}

This work is a continuation of the work~\cite{I} where some
eigenstates were introduced for a model based on the solutions
to the tetrahedron equations described in paper~\cite{k2}.
It contains two separate observations on two sorts
of eigenstates introduced in~\cite{I}: particle-like states
(plane waves and their superpositions), and string-like states.
Those observations are explained in Sections \ref{sec-alg-one}
and~\ref{sec-string-sym} respectively. The reasons for doing
this rather technical work are explained in Section~\ref{II-discussion}.

Let us give some definitions and remarks. We will depict
the operators graphically in such a way that each operator
will have some number of ``incoming edges'' and the same number
of ``outgoing edges'' (or ``links''). To each edge corresponds
its own copy of a two-dimensional complex linear space,
and to several edges of the same (incoming or outgoing) kind
together corresponds the tensor product of their spaces.
Each of the mentioned two-dimensional spaces has a basis
of a {\em 0-particle\/} and {\em 1-particle\/} vectors.

For any, maybe infinite, collection of edges, we will define
this {\em collection's 0-particle vector}, or {\em vacuum}, as
the tensor product of 0-particle vectors throughout the collection
(in this paper, the meaning of infinite tensor products will
be always clear). Further, we will identify
a 1-particle vector in an edges with its tensor product with
the 0-particle vectors in all the collection's other edges and
define a {\em collection's 1-particle vector\/} as a formal sum
over all its edges of the corresponding 1-particle vectors,
with any complex coefficients. Then, we can define in an obvious
way the 2-particle, 3-particle etc.\ states.

According to the above, an operator acts from
the tensor product of ``incoming'' spaces to the tensor product
of ``outgoing'', i.e.\ different, spaces.
Still, sometimes we will assume that all the
edges along one straight line represent {\em the same\/}
two-dimensional space. This is convenient e.g.\ when we write
out the tetrahedron equation, as in formula~(\ref{tetr}) below,
and this will never lead to confusion.

\section{Algebraization of one-particle-state construction}
\label{sec-alg-one}

In this section, we will present a one-parameter family of
``creation operators''. When applied to the ``vacuum'', these
operators produce one-particle states---plane waves, described already
in the paper~\cite{I}. As we will see, the very construction of these
operators presupposes that they act on vectors which don't differ
much from the ``vacuum''. We will not try to make this statement
more exact here. Instead, in this paper we will assume from the beginning
that the domain of definition of those operators consists of only
one-dimensional space generated by vacuum, leaving the extension
of that domain for further work.

\subsection{Description of transfer matrices from which the creation
 operators are constructed}
\label{subsec-1}

Creation operators will be transfer matrices on a kagome lattice
with some special boundary conditions. Graphically, such a transfer
matrix is depicted in Figure~\ref{figII-1}.
\begin{figure}[hbtp]
\begin{center}
\unitlength=1.00mm
\linethickness{0.4pt}
\begin{picture}(120.00,50.00)
\put(25.00,0.00){\line(1,1){50.00}}
\put(45.00,0.00){\line(1,1){50.00}}
\put(65.00,0.00){\line(1,1){50.00}}
\put(35.00,15.00){\line(1,0){50.00}}
\put(55.00,35.00){\line(1,0){50.00}}
\put(20.00,0.00){\line(2,1){100.00}}
\put(40.00,0.00){\line(2,1){60.00}}
\put(60.00,0.00){\line(2,1){20.00}}
\put(40.00,20.00){\line(2,1){60.00}}
\put(60.00,40.00){\line(2,1){20.00}}
\put(30.00,2.00){\line(0,1){6.00}}
\put(50.00,2.00){\line(0,1){6.00}}
\put(70.00,2.00){\line(0,1){6.00}}
\put(40.00,12.00){\line(0,1){6.00}}
\put(50.00,12.00){\line(0,1){6.00}}
\put(60.00,12.00){\line(0,1){6.00}}
\put(70.00,12.00){\line(0,1){6.00}}
\put(80.00,12.00){\line(0,1){6.00}}
\put(50.00,22.00){\line(0,1){6.00}}
\put(70.00,22.00){\line(0,1){6.00}}
\put(90.00,22.00){\line(0,1){6.00}}
\put(60.00,32.00){\line(0,1){6.00}}
\put(70.00,32.00){\line(0,1){6.00}}
\put(80.00,32.00){\line(0,1){6.00}}
\put(90.00,32.00){\line(0,1){6.00}}
\put(100.00,32.00){\line(0,1){6.00}}
\put(70.00,42.00){\line(0,1){6.00}}
\put(90.00,42.00){\line(0,1){6.00}}
\put(110.00,42.00){\line(0,1){6.00}}
\end{picture}
\end{center}
\caption{}
\label{figII-1}
\end{figure}
We are considering the eigenstates of transfer matrix
made up of ``hedgehogs'', as in work~\cite{I}. Naturally, a
kagome transfer matrix must be such that
it would be possible to bring the hedgehogs through it using
the tetrahedron equation.

The present work, as well as~\cite{I},
is based only on simple solutions of that equation that
can be found in~\cite{k2}. For those
solutions, the following form of equation holds:
\be
\hbox{$\matrix{
S_{01,02,12}(\varphi_0,\varphi_1,\varphi_2)
S_{01,03,13}(\varphi_0,\varphi_1,\varphi_3)
S_{02,03,23}(\varphi_0,\varphi_2,\varphi_3)
S_{12,13,23}(\varphi_1,\varphi_2,\varphi_3)
\hfill \vadjust{\smallskip} \cr
 =
S_{12,13,23}(\varphi_1,\varphi_2,\varphi_3)
S_{02,03,23}(\varphi_0,\varphi_2,\varphi_3)
S_{01,03,13}(\varphi_0,\varphi_1,\varphi_3)
S_{01,02,12}(\varphi_0,\varphi_1,\varphi_2).
\hfill} $}
\label{tetr}
\ee
Here a number 0, 1, 2, or 3 is attached to a {\em plane}, that is,
to a face of the tetrahedron. An operator $S_{ij,ik,jk}$ acts
in the tensor product of three linear spaces corresponding
to the {\em lines}---intersections of those planes.

Assuming that parameters $\varphi_1,\varphi_2,\varphi_3$ belong
to the ``hedgehogs'' $S_{12,13,23}$ and are given, there remains one free
parameter~$\varphi_0$, where the number 0 is attached to the plane
of the kagome lattice.

\subsection{Boundary conditions for creation operators}

It will take some effort to describe the boundary conditions that we
are going to impose on kagome transfer matrices of
Subsection~\ref{subsec-1} to obtain out of them creation operators.
The problem is that we are considering an {\em infinite\/} in all
plane directions kagome lattice. So, first, let us draw
in Figure~\ref{figII-2}
\begin{figure}[hbtp]
\begin{center}
\unitlength=1.00mm
\linethickness{0.4pt}
\begin{picture}(120.00,120.00)
\thicklines
\put(0.00,10.00){\line(1,0){89.00}}
\put(91.00,10.00){\line(1,0){18.00}}
\put(111.00,10.00){\line(1,0){9.00}}
\put(0.00,50.00){\line(1,0){49.00}}
\put(51.00,50.00){\line(1,0){18.00}}
\put(69.00,50.00){\line(0,0){0.00}}
\put(71.00,50.00){\line(1,0){49.00}}
\put(0.00,90.00){\line(1,0){9.00}}
\put(11.00,90.00){\line(1,0){18.00}}
\put(31.00,90.00){\line(1,0){89.00}}
\put(30.00,0.00){\line(0,1){69.00}}
\put(30.00,71.00){\line(0,1){18.00}}
\put(30.00,91.00){\line(0,1){29.00}}
\put(70.00,0.00){\line(0,1){29.00}}
\put(70.00,31.00){\line(0,1){18.00}}
\put(70.00,51.00){\line(0,1){69.00}}
\put(110.00,0.00){\line(0,1){9.00}}
\put(110.00,11.00){\line(0,1){109.00}}
\put(0.00,80.00){\line(1,1){9.00}}
\put(11.00,91.00){\line(1,1){29.00}}
\put(0.00,40.00){\line(1,1){29.00}}
\put(31.00,71.00){\line(1,1){49.00}}
\put(0.00,0.00){\line(1,1){49.00}}
\put(51.00,51.00){\line(1,1){69.00}}
\put(40.00,0.00){\line(1,1){29.00}}
\put(71.00,31.00){\line(1,1){49.00}}
\put(80.00,0.00){\line(1,1){9.00}}
\put(91.00,11.00){\line(1,1){29.00}}
\put(10.00,90.00){\circle{2.00}}
\put(30.00,90.00){\circle{2.00}}
\put(30.00,70.00){\circle{2.00}}
\put(50.00,50.00){\circle{2.00}}
\put(70.00,50.00){\circle{2.00}}
\put(70.00,30.00){\circle{2.00}}
\put(90.00,10.00){\circle{2.00}}
\put(110.00,10.00){\circle{2.00}}
\put(10.00,90.00){\makebox(0,0)[cc]{.}}
\put(30.00,90.00){\makebox(0,0)[cc]{.}}
\put(30.00,70.00){\makebox(0,0)[cc]{.}}
\put(50.00,50.00){\makebox(0,0)[cc]{.}}
\put(70.00,50.00){\makebox(0,0)[cc]{.}}
\put(70.00,30.00){\makebox(0,0)[cc]{.}}
\put(90.00,10.00){\makebox(0,0)[cc]{.}}
\put(110.00,10.00){\makebox(0,0)[cc]{.}}
\put(-2.00,93.00){\makebox(0,0)[lb]{$B$}}
\put(22.00,108.00){\makebox(0,0)[lb]{$D$}}
\put(113.00,18.00){\makebox(0,0)[lb]{$C$}}
\put(88.00,3.00){\makebox(0,0)[lb]{$A$}}
\put(30.00,96.00){\vector(0,1){0.00}}
\put(5.00,90.00){\vector(1,0){0.00}}
\put(8.00,88.00){\vector(1,1){0.00}}
\put(28.00,68.00){\vector(1,1){0.00}}
\put(30.00,65.00){\vector(0,1){0.00}}
\put(45.00,50.00){\vector(1,0){0.00}}
\put(48.00,48.00){\vector(1,1){0.00}}
\put(68.00,28.00){\vector(1,1){0.00}}
\put(70.00,25.00){\vector(0,1){0.00}}
\put(85.00,10.00){\vector(1,0){0.00}}
\put(88.00,8.00){\vector(1,1){0.00}}
\put(36.00,90.00){\vector(1,0){0.00}}
\put(70.00,56.00){\vector(0,1){0.00}}
\put(76.00,50.00){\vector(1,0){0.00}}
\put(110.00,16.00){\vector(0,1){0.00}}
\put(103.00,10.00){\vector(1,0){0.00}}
\put(100.00,20.00){\vector(1,1){0.00}}
\put(80.00,40.00){\vector(1,1){0.00}}
\put(70.00,43.00){\vector(0,1){0.00}}
\put(63.00,50.00){\vector(1,0){0.00}}
\put(60.00,60.00){\vector(1,1){0.00}}
\put(40.00,80.00){\vector(1,1){0.00}}
\put(30.00,83.00){\vector(0,1){0.00}}
\put(23.00,90.00){\vector(1,0){0.00}}
\put(20.00,100.00){\vector(1,1){0.00}}
\thinlines
\multiput(87.00,3.00)(-5.00,5.00){18}{\line(-1,1){4.00}}
\multiput(112.00,18.00)(-5.00,5.00){18}{\line(-1,1){4.00}}
\end{picture}
\end{center}
\caption{}
\label{figII-2}
\end{figure}
the lattice viewed from above
(here, we have deformed the lattice a bit in comparison
with the paper~\cite{I}).
Then let us draw a dashed line $AB$ and cut off for a while
the part of the lattice lying to the left of that line
(it will be explained in Subsection~\ref{subsec-calc}
that really there is much arbitrariness in choosing the line $AB$,
but let it be for now as in Figure~\ref{figII-2}). For the rest
of transfer matrix, let us define the boundary condition along $AB$
as follows. Consider all the lattice edges intersecting $AB$.
They are {\em incoming edges\/} for the remaining part of transfer
matrix. To define the boundary conditions, we must indicate some vector
$\Sigma_{AB}$ in the tensor product of corresponding spaces. Let us assume
that $\Sigma_{AB}$ is a {\em 1-particle vector\/} as defined
in the Introduction,
whose exact form is to be determined.

Consider the band---the part of transfer matrix lying between
the lines $AB$ and $CD$. This band represents an operator acting from
the space corresponding to its incoming edges into the space
corresponding to its outgoing edges, where the incoming edges
are those intersecting $AB$ and those pointing from behind the kagome
lattice plane into the vertices situated within the band,
while the outgoing edges are those intersecting $CD$
and those pointing from the vertices situated within the band
at the reader. The latter vertices
are marked $\alh$ in Figure~\ref{figII-2}.

Let us require that our band operator---let us call it $\cal B$---transform
the tensor product
$\Sigma_{AB} \otimes \Omega_{\alh}$, where $\Omega_{\alh}$
is the vacuum for the set of edges pointing into the ``$\alh$'' vertices,
into the following sum:
\be
{\cal B}\, \Sigma_{AB} \otimes \Omega_{\alh} =
\kappa\,\Sigma_{CD} \otimes \Omega'_{\alh} + \Omega_{CD}\otimes
\Psi_{\alh},
\label{calB}
\ee
where $\kappa$ is a number; $\Sigma_{CD}$ is the vector similar to
$\Sigma_{AB}$, but corresponding to the edges situated one lattice
period to the right, i.e.\ intersecting $CD$; $\Omega'_{\alh}$ is
the vacuum for edges pointing from the ``$\alh$'' vertices to the reader
(who can thus identify $\Omega'_{\alh}$ with $\Omega_{\alh}$ if desired);
$\Omega_{CD}$ is the vacuum for the set of vectors intersecting $CD$;
$\Psi_{\alh}$ is some vector lying in the same tensor product of
spaces as $\Omega'_{\alh}$.
It is remarkable that, for any $\varphi_0$, relation (\ref{calB})
can be satisfied
with a proper choice of~$\Sigma_{AB}$. The vector $\Psi_{\alh}$ will then
be a 1-particle vector. Some details of calculations concerning
relation~(\ref{calB}) are explained in Subsection~\ref{subsec-calc},
while here we are going to {\em use\/} this relation.

The next band, lying to the right of $CD$, has vector
$\kappa\,\Sigma_{CD}$
as its incoming vector. Thus, the whole situation is repeated up to
the factor~$\kappa$. On the other hand, we could have cut the lattice,
instead of the line $AB$, along some other line lying, e.g.,
$n$ lattice periods to the left. In that case, we should have taken
for the incoming vector the vector $\Sigma_{AB}$ shifted by $n$
periods to the left and multiplied by~$\kappa^{-n}$.
Letting $n\to\infty$, we get a $\Psi_{\alh}$-like vector
in {\em every\/} band of the sort depicted in Figure~\ref{figII-2}.
Summing up
all those vectors, we get a 1-particle state of the same kind as in
paper~\cite{I}. Those states are now parameterized by
the parameter~$\varphi_0$.

In fact, one more boundary condition must be imposed at the
``right infinity'' of the lattice. This is explained in the end of
Subsection~\ref{subsec-calc}.

\subsection{Some technical details}
\label{subsec-calc}

Instead of the straight line $AB$ in Figure~\ref{figII-2}, we could use
any (connected) curve $l$ {\em intersecting each straight line of the kagome
lattice exactly one time\/} in such way that a boundary condition
is given in the tensor product corresponding to the edges that
intersect~$l$. Assuming that a 1-particle vector is given as the boundary
condition, we will require that after any deformation of~$l$ such that
it passes through {\em one\/} of the lattice vertices, the boundary
condition remain to be 1-particle.

In other words, the mentioned vertex is added to or withdrawn from
the considered part of the lattice. Let that vertex be, e.g., such as
in Figure~\ref{figII-3}.
\begin{figure}[hbtp]
\begin{center}
\unitlength=1mm
\linethickness{0.4pt}
\begin{picture}(40.00,40.00)
\put(0.00,0.00){\line(1,1){19.00}}
\put(21.00,21.00){\line(1,1){19.00}}
\put(0.00,20.00){\line(1,0){19.00}}
\put(21.00,20.00){\line(1,0){19.00}}
\put(0.00,0.00){\vector(1,1){12.00}}
\put(0.00,20.00){\vector(1,0){12.00}}
\put(23.00,20.00){\vector(1,0){7.00}}
\put(23.00,23.00){\vector(1,1){7.00}}
\put(20.00,20.00){\circle{2.00}}
\put(20.00,20.00){\makebox(0,0)[cc]{.}}
\put(3.00,21.00){\makebox(0,0)[cb]{$a$}}
\put(3.00,5.00){\makebox(0,0)[cb]{$b$}}
\put(35.00,22.00){\makebox(0,0)[cb]{$f$}}
\put(35.00,37.00){\makebox(0,0)[cb]{$c$}}
\end{picture}
\end{center}
\caption{}
\label{figII-3}
\end{figure}
An incoming 1-particle vector for it is described
by two amplitudes $a$ and $b$, with $a$ corresponding to the edge~$01$
(see the text just after equation~(\ref{tetr})) and $b$---to the edge~$02$.
It is required that the result of transforming this incoming vector
by the matrix~$S^{\rm T}$ (we assume, as in work~\cite{I}, that in each
vertex there is a matrix of the same type as on page~96 of paper~\cite{k2},
but transposed) contain no three-particle part. This leads at once
to the condition
\be
{a\over b}=\,\root\of{\tanh(\varphi_0-\varphi_1)}
 \,\,\,\root\of{\tanh(\varphi_0-\varphi_2)}.
\ee

The other ratios of the amplitudes written out in Figure~\ref{figII-3}
are simply the matrix elements of~$S^{\rm T}$:
\be
{c\over a}={f\over b}=\,\root\of{\tanh(\varphi_0-\varphi_1)}
 \,\,\,\root\of{\cotanh(\varphi_0-\varphi_2)}.
\ee
Similar relations can be written for the kagome lattice vertices
of two other kinds, that is
$\matrix{
\unitlength=1mm
\linethickness{0.4pt}
\begin{picture}(6.00,7.00)
\put(0.00,3.50){\line(1,0){6.00}}
\put(3.00,0.50){\line(0,1){6.00}}
\end{picture} }$
and
$\matrix{
\unitlength=1mm
\linethickness{0.4pt}
\begin{picture}(6.00,7.00)
\put(3.00,0.50){\line(0,1){6.00}}
\put(0.00,0.50){\line(1,1){6.00}}
\end{picture} }$.
It turns out that all those relations together determine the amplitudes
at {\em all\/} lattice edges from a given one of them without contradiction.
Thus, the amplitudes for the $\alh$-edges, that are incoming for
the hedgehog transfer matrix, are determined correctly.

Those amplitudes give exactly its eigenstate for any fixed~$\varphi_0$.
This can be proved by a rather obvious reasoning:
use the tetrahedron equation and the possibility to express the amplitudes
at different edges through one another. The details are left
for the reader.

There remains, however, another detail that is important:
we must impose one more boundary condition, that is at the
``right infinity'' (see again Figure~\ref{figII-2}).
In order to obtain the 1-particle vector at the $\alh$-edges,
and no {\em vacuum\/} component, let us take a straight
line $C'D'$---like $CD$, but somewhere far to the right---and
require any 1-particle vector in the space corresponding
to edges that intersect $C'D'$ be multiplied by zero,
while the vacuum vector in that space be left intact.
This can be interpreted as taking the scalar product of the vector
at the edges intersecting $C'D'$ and the ``vacuum covector''.
Then, of course, we let $C'D'$ tend to the right infinity,
so all this procedure does not change the 1-particle component
of the $\alh$-vector, but the vacuum component vanishes.

\section{Strings as symmetries}
\label{sec-string-sym}

Let us introduce the Pauli matrices
$$
\sigma_1=\pmatrix{0 & 1 \cr 1 & 0}, \qquad
\sigma_2=\pmatrix{0 & -i \cr i & 0}, \qquad
\sigma_3=\pmatrix{1 & 0 \cr 0 & -1},
$$
as well as a unity matrix
$$
\sigma_0=\pmatrix{1 & 0 \cr 0 & 1}.
$$
Note that the subscripts of these matrices have other meaning than
the subscripts of $S$-matrices in equations like~(\ref{tetr}).

It follows from the explicit form of $S$-matrices given in paper~\cite{k2}
that the $S$-matrices commute with the operators
\be
\sigma_2 \otimes \sigma_2 \otimes \sigma_0, \quad
\sigma_0 \otimes \sigma_1 \otimes \sigma_1 \hbox{ and }
\sigma_1 \otimes \sigma_0 \otimes \sigma_2.
\label{II-*}
\ee

If now we select some set of the kagome lattice
horizontal lines (to be exact, of those
depicted in Figure~\ref{figII-2} as horizontal)
and consider the tensor product of matrices $\sigma_2$ over all
vertices belonging to those lines, then the hedgehog transfer matrix~$T$
will be permutable with that product up to the fact that the lines move
in the lattice plane, as explained in work~\cite{I}
(the lines result from
the intersection of cubic lattice faces with a plane perpendicular
to a cube's spatial diagonal, and move in that plane when the plane itself
moves). This permutability follows
immediately from the fact that $S$ commutes with the first
of operators~(\ref{II-*}).
Similarly, it is not difficult to formulate the analogous statements
for sets of oblique and vertical lines, using respectively the second
and third of products~(\ref{II-*}).

Using the described symmetries of transfer matrix~$T$, we can,
starting from any state vector~$\Theta_0$ whose evolution under
the action of degrees of~$T$ we can describe in this or that way
(recall that $T$
is such that its degrees are represented graphically as
``oblique layers'' of the cubic lattice), build many new states~$\Theta$
whose evolution we will also be able to describe.

\section{Discussion}
\label{II-discussion}

The aim of our ``algebraization'' is, of course, to learn to construct
multi-particle states for $2+1$-dimensional models. Let us remind that
only some special two-particle states have been constructed
in paper~\cite{I}, and even the superposition of two arbitrary
one-particle states from that very work has not been obtained
there.

In this paper, we still do not present the construction of multi-particle
states. The reason for doing this work is a hope that,
with the help of a proper ``regularization'', the action of our
creation operators can be extended from just vacuum onto
one-particle, two-particle and so on (eigen)vectors.

It is clear that the superposition of two arbitrary one-particle
eigenstates from work~\cite{I} cannot lie just in the 2-particle
space as it is defined in Introduction. So, probably, some new
terminology should be introduced to distinguish between the
2-{\em particle\/} space and 2-{\em excitation\/} states.

As for the strings providing symmetries and thus multiplying
the eigenstates, it is still to be clarified whether those strings are
particular cases of a family including more interesting species.

Finally, it is certainly interesting to find eigenstates for the model
based on other simple solutions to the tetrahedron equation~\cite{h},
and perhaps for the general model described in~\cite{mss}.

\end{document}